# On the true nature of transfer reactions leading to the complete fusion of projectile and target

## G. Mouze and C. Ythier

Université de Nice , Faculté des Sciences, 06 108 Nice cedex 2, France.

mouze@unice.fr

**Abstract**

The transfer of nucleons in hot-fusion reactions occurs within 0.17 yoctosecond , in a new state of nuclear matter. We suggest that the same state should show itself in an early stage of the phenomena occurring in nucleus-nucleus collisions realized at relativistic energies.

PACS 25.70 Hi – Transfer reactions

PACS 25.70 Jj – Fusion and fusion-fission reactions

PACS  25.75. Ld – Collective flow

**1. Introduction**

In a recent communication devoted to the attempts made in the eighties by nuclear chemists to create heavy nuclei by transfer of heavy ions, in the hope of creating new heavy elements [1], we showed that a great resemblance exists between isotopic distributions encountered in cold-fission experiments [2,3], on the one hand, and in transfer reactions leading to trans-target nuclei [4-9], on the other hand; and we suggested that this resemblance proves that one and the same state of nuclear matter intervenes in both cases. This state, referred to as "nucleon phase" [10] ,has a so short lifetime, of only 0.17 yoctosecond (1 ys = $10^{-24}$ s) [11], that uncertainties in A, in Z, and in N remain attached to the final product. And we briefly pointed out that even transfer reactions leading to a complete fusion of projectile and target − such as those now used for the synthesis of superheavy elements, e.g. hot-fusion reactions [12,13]− show the same surprising diversity of reaction products, due to the same uncertainties in A, Z and N.

The present communication should have been devoted to a tentative answer to the question:

"Could the existence of this new state of nuclear matter, put into evidence not only in fission but also in transfer reactions, have an impact on the investigation of *the initial phenomena occurring in A/A collisions* realized at relativistic energies?"

In fact, this answer will be limited, 1°) to a more detailed study of the hot-fusion reaction, and 2°) to a first comparison of these initial phenomena with the hot fusion (Sects.2 and 3).And since many properties of the fission reaction, up to now not observed in the hot-fusion reaction, might nevertheless throw a new light on these phenomena, Sect.4 will be devoted to a recalling of the most intriguing properties of the fission reaction.

## 2. An example of transfer reaction leading to the complete fusion of projectile and target.

There are two kinds of fusion reactions, but they differ only in the choice of projectiles and targets. The "cold fusion" reaction [14] uses $^{208}$Pb or $^{209}$Bi targets and a massive projectile ( $A \geq 50$), whereas the "hot fusion" reaction [15] uses $^{244}$Pu, $^{248}$Cm or $^{249}$Cf targets and $^{48}$Ca projectiles. It is why we limit our discussion to the hot fusion.

As an example of hot fusion reaction, let us consider the process:

(1) $^{48}$Ca + $^{244}$Pu → $^{292}$(114)

leading to the formation of *several* flerovium isotopes [13].

Indeed, precisely because this reaction is a transfer reaction, it cannot lead to a single product, but only to an "isotopic distribution", as shown in ref.[1].

More precisely, this reaction cannot, at the same time, abide by the law of matter conservation, clearly respected in eq.(1), and the energy-time uncertainty relation, which plays a major role in any transfer reaction. Indeed, due to the uncertainty in the neutron number N imposed to the transfer product by the extreme brevity of the intermediary new state, a number of "prompt "neutrons must be emitted in the formation of the most probable transfer product [1,11]: On the contrary, the process described by eq.(1) clearly corresponds to the emission of zero neutron, as required by matter conservation.

Let us show that the data reported by the authors of ref.[13] precisely suggest an "isotopic distribution" represented by a Gaussian curve having a full width at half maximum (fwhm) of about 2.54 atomic mass units (u), as expected for a transfer reaction:

According to Oganessian et al. [13], the greatest yield of the fission reaction has been observed in the configurations given by the following eqs. (2) and (3):

(2) 243MeV – $^{48}$Ca + $^{244}$Pu

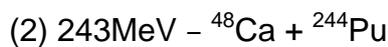

$\rightarrow {}^{289}F\ell + 3n$ (2 events)

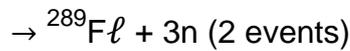

$\rightarrow {}^{288}F\ell + 4n$ (7 events)

(3) 250MeV – $^{48}$Ca + $^{244}$Pu

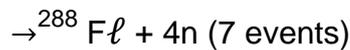

$\rightarrow {}^{289}F\ell + 3n$ (1 event)

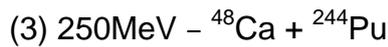

$\rightarrow {}^{288}F\ell + 4n$ (4 events)

Eq.(3) and eq.(1) suggest a Gaussian yield curve, which is centered on 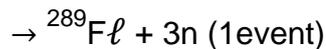 = 288.20, i.e. on the neutron number 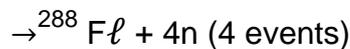 = 174.20 as shown in fig.1. Eqs. (2) and (1) would lead to almost the same yield curve, but centered on  = 288.22 (i.e on  = 174.22).

One sees that at $E_{beam}$ = 250 MeV the most probable number of emitted prompt neutrons should be  = 3.80, but would be  = 3.78 for the data corresponding to $E_{beam}$ = 243MeV.

Fig.1 allows the following conclusion: The hot fusion reaction leading to the flerovium isotopic distribution represented in this figure is a transfer reaction occurring within a new state of nuclear matter which has a lifetime of about 0.17 ys, and which cannot be distinguished from the state occurring in the rearrangement step of nuclear fission and in the transfer reactions leading to trans-target isotopes as reported in ref.[1].

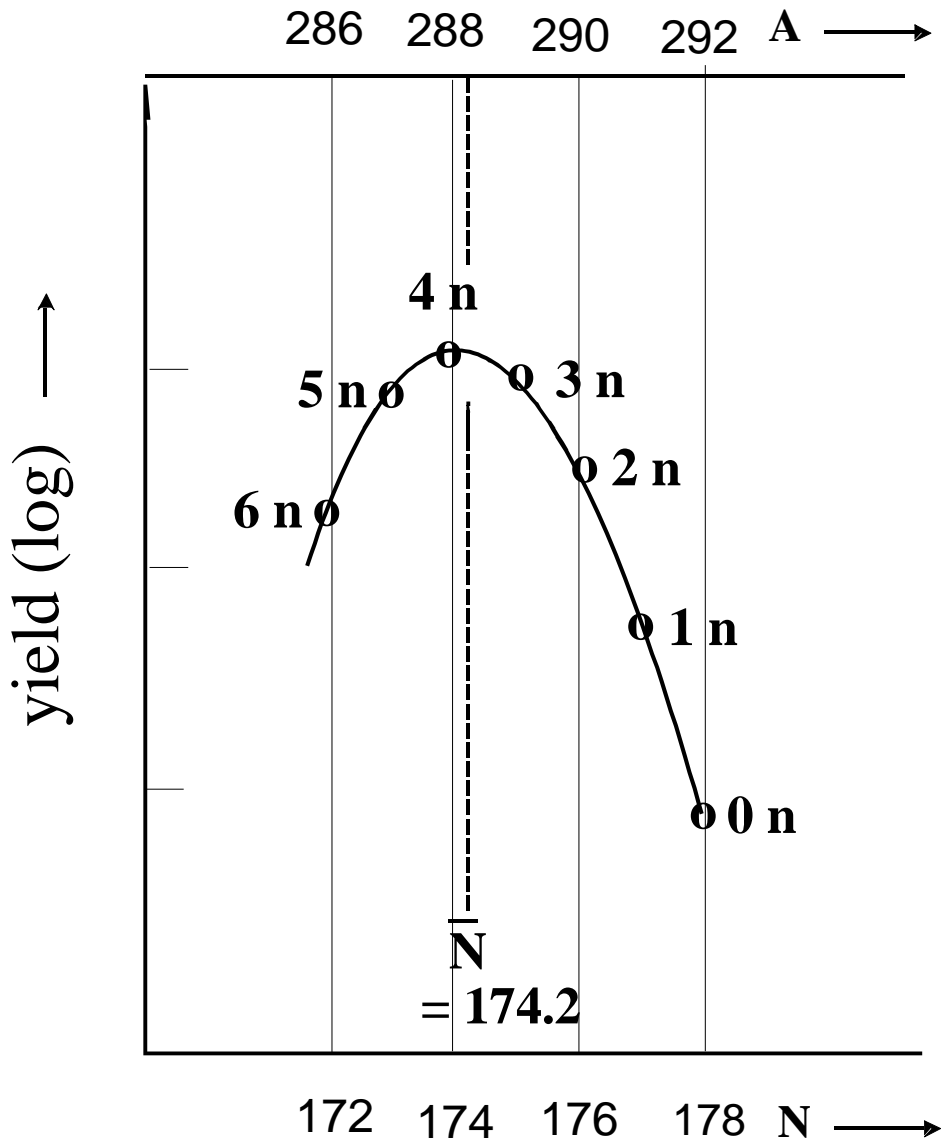

Fig.1: Suggested yield curve for the flerovium isotopes (Z=114) formed in the collision $^{48}$Ca + $^{244}$Pu at $E_{beam}$ = 250 MeV

As expected, the width of this yield curve is really about 2.14 u, a measure of the uncertainty in N resulting from the uncertainty law.

This yield curve proves the great diversity of the products of the hot fusion reaction represented by eq.(1). Indeed, it shows, for example, that the expected yield of the isotope $^{287}$F$\ell$ – which has been observed only at 257 MeV, and only in one event, according to refs. [12,13]– is in fact almost as great as the reported yield of $^{289}$F$\ell$.

In fact, this diversity is even greater than shown by fig.1, because, as a consequence of the uncertainty in Z (of the order of 1.637 u [11]), the competition of products with Z = 113 and Z = 115 cannot be excluded. Even more, this diversity should appear too in the spontaneous-fission phenomena accompanying the alpha-decay chains of the fusion products. And this diversity might result not only from the fission of extra "heavy target-like transfer products" – as recently suspected by Heinz et al. [16] – but also from the hot fusion process itself, due to its complex isotopic distribution.

The name of evaporation residue, up to now given to the fusion product, is not quite appropriate, since it suggests that the neutron emission results from an evaporation caused by an energy increase, whereas it is caused by the lifetime of only 0.17 ys of the nucleon phase.

### 3. Comparison of A/A collisions with the hot fusion

"Can a resemblance exist between the transfer reactions occurring in lead/lead, or gold/gold, or even p/p collisions realized at relativistic energies, and those occurring in fusion reactions such as those used for the synthesis of superheavy elements?"

This question may at first sight appear too daring, because the intriguing collective flow [17] which develops in what has been called a quark-gluon plasma [18] is now investigated in several places at collision energies going from 1 A GeV to 2.76 A TeV [19,20]; on the contrary, the synthesis of superheavy nuclei realized e.g. at Dubna does not require beam energies exceeding one GeV.

However, if the lifetime of the new state occurring in transfer reaction is considered as an uncertainty in time, then, according to the energy-time uncertainty relation, to this Δt corresponds a possible energy-release of up to 3.87 GeV. And this energy, does not notably differ from the energy-density postulated for the "fireball", namely ~ 3.85 GeV according to the authors of ref.[19]. It is a first point.

Moreover, important "fluctuations" have been reported to exist in the dense hadronic matter of the fireball [20,21]. But these fluctuations might be related to the uncertainties in A, Z and N resulting from the brevity of a transfer reaction and amounting to 4.175 u for ΔA, to about 1.64 u for ΔZ and to about 2.54 for ΔN [11]. It is a second point.

And the threshold energy, where a quark-gluon plasma might just be created, is perhaps situated at lower energies.

In fact, new colliders, which aim at reaching this threshold, are now under construction [22]. It is a third point.

**4. Comparison with the reaction of fission**.

It may be asked too whether properties of the nucleon phase other than those quoted above have been observed in the fission reaction and might play a role in the medium of the fireball.

*4-A* The first of these properties is *the character particularly "easy" of the nucleon transfer* within the nucleon phase. Indeed, it can be shown that part of the nucleons released in the destruction of the $^{208}$Pb core present in any fissioning system are transferred to the primordial cluster *within a time that might be even shorter than* 0.17 ys.

For example, in the neutron-induced fission of $^{239}$Pu, part of the 82 nucleons released in the process

(4) $^{32}$Mg + $^{208}$Pb → $^{32}$Mg + A = 126 nucleon core + 82 free nucleons

are transferred so rapidly to the $^{32}$Mg cluster and form with it so rapidly an A = 82 nucleon core, according to

(5) $^{32}$Mg + 82 free nucleons → A = 82 nucleon core + (82-32) free nucleons

that the light-product *yield*, in the region of "far-asymmetric fission" [23] extending from the A-value of the cluster, i.e. A = 32, to A = 82, is *extremely low*. And an exceptionally simple correlation between the logarithm of the yield and the fission energy has been found to hold in this region [24].

*4-B* The second of these properties is *the apparent disappearance of any proton charge during 0.17 ys.*

This situation is clearly demonstrated by the creation of the two nucleon cores made of 126 and 82 nucleons: In the rearrangement step of the fission reaction, the neutron shells and the proton shells, which coexist in ordinary nuclear matter, suddenly disappear and are replaced by a new kind of shells, which are closed at the two magic mass numbers 126 and 82. And such a situation necessarily results from the disappearance of any proton charge, since the only difference between the "proton phase" and the "neutron phase " of ordinary nuclear matter was the presence, in the first one, of an additional potential to that resulting from the spin-orbit coupling [25]. This additional potential is precisely the Coulomb potential which is due to these proton charges. It is why the medium constituting the "new phase" seems to be made of "neutrons" but has been provisionally called "nucleon phase"[10], the word nucleon being there taken in the sense once given by W. Heisenberg to a particle which can become either a proton or a neutron [26].

It may be asked whether this disappearance of any proton charge is *responsible* for the easy nucleon transfer.

In addition to the theoretical arguments quoted above, there exists an experimental fact requiring the existence of *a medium exclusively made of "nucleons":* It is the law of K.F. Flynn et al. [27], saying that *the mean mass of the light fission product increases linearly with the mass of the fissioning nucleus.* And this situation can be explained only if one assumes that the *formation of any light fission product requires the transfer of an, on average, constant number of nucleons* from the $^{208}$Pb primordial core to the cluster. This number is 68 [23] –if the emission of prompt neutrons is neglected–: It means that the medium constituting the nucleon phase is exclusively made of the transferred nucleons, and of hard cores made of nucleons. Any charge is absent – or masked–.

Let us recall that this "nucleon phase" model explains not only the mass distribution of the asymmetric fission mode and of the symmetric fission mode [28,29] , but also the rare "collinear ternary" fission mode recently discovered by Yu.V. Pyatkov et al. [30,31].

*4-C* The properties reported in the preceding Sects. 4A and 4B can be summarized like this: the nucleon phase is essentially the speedy setting up of an overall structure. But, in these conditions, *how can the astonishing diversity of the fission products, in particular their various Z- values, be created* ?

We have already shown in Sect.2 how the emission of prompt neutrons allows the creation of isotopic distributions centered on the most probable value   of N. But the creation of products having different values of Z , distributed around the most probable  value  of the atomic number, should require the emission of *prompt protons*. In a nucleon phase, where any proton charge has disappeared, and which is made of nucleons, i.e. of virtual neutrons or protons, it is reasonable to assume that the prompt protons are, in reality, emitted *as prompt neutrons*, and that the proton charge of the products appears only at the end of the nucleon phase. Thus *the measured prompt neutron yield* should represent the sum of that corresponding to the formation of isotopes and of that corresponding to the formation of isotones…

## 5. Conclusion

We have shown that the isotopic distributions created in hot fusion reactions are characterized by the same width at half maximum than that observed in all distributions of isotopic trans- target products of transfer reactions. It means that *the hot fusion products are formed in the same new state of nuclear matter, having a lifetime of only 0.17 ys.*

It is tempting to conclude that the same state of nuclear matter of 0.17 ys is essentially created in the initial state of lead/lead or gold/gold or even proton/ proton collisions at relativistic energies, because these collisions are transfer reactions leading to a complete fusion. The energy density of about 3.85 GeV reported for the fireball [19] corresponds almost exactly to that of the nucleon phase. *The lifetime of the fireball should also be of  ~ 0.17 ys*  And the dense hadronic matter created in these collisions should present *density fluctuations* equal to the expected uncertainties in the A-, Z- and N- numbers.

But the study of the fireball might furnish new information on the properties of this new state of nuclear matter, for example on the intriguing disappearance of the proton charge in the nucleon phase. Might a collective interaction with $W^+$ and $W^-$ boson fields intervene [32] ? For all these reasons a new investigation of the initial state [22] of the fireball is desirable.

The dominant role played in the nucleon phase by the energy-time uncertainty relation raises several questions. For example, is it satisfactory to refer to a relation which is itself seriously considered as "not firmly founded"?

Anyway, new keys have to be found for a deeper investigation of the true nature of the fireball at relativistic energies.

It will be the aim of our next communication.

**References**

[1] G.Mouze, S. Hachem and C. Ythier, http://arxiv.org/abs/1204.2647 [nucl-exp] april12, 2012.

[2] C. Signarbieux, *1st Intern. Conf. on Dynamical Aspects of Nuclear Fission*, Smolenice, Slovakia, J. Kristiak and B.I.Pustylnik, eds., J.I.N.R.,Dubna,1992,p.19.

[3] G.Mouze and C. Ythier, *Nuovo Cimento A* **106** (1993) 835.

[4] A.G. Demin et al. *Intern. Symp. On the Synthesis and Properties of New Elements*, Dubna, 1980, abstract D-80-556, p.60.

[5] D.Lee et al.,*PRC* ,**25** (1982)286.

[6] D.C. Hoffman et al.,*PRC* **31** (1985) 1763.

[7] M. Schädel et al.,*PRC* **33** (1986) 1547.

[8] H.V. Scherrer et al., *Z. Phys A* **335** (1990) 421.

[9] Gäggeler et al. *PRC* **33** (1986) 1983.

[10] G.Mouze and C. Ythier, *46th intern.Winter Meeting on Nuclear Physics*, Bormio, Italy, Jan.20-26,2008, I. Iori and A. Tarantola,eds.,Università degli Studi di Milano,2008,p.230.

[11] G. Mouze and C. Ythier, http://arxiv.org/abs/1004.1337  [nucl-exp] april 8, 2010.

[12] Yu. Ts. Oganessian, *J.Phys.* G **34**, (2007) R165.

[13] Yu. Ts. Oganessian et al. *PRC* **69** (2004) 054607.

[14] Yu. Ts. Oganessian, *Lecture Notes in Physics*, vol.33, J. Echlers, ed., Springer 1975,p.222.

[15] Yu. Ts. Oganessian et al.,*PRC* **62** (2000) 041604.

[16] S. Heinz et al.,*EPJA* **48** (2012) 32.

[17] see, e.g. K. Aasmodt et al. (ALICE Coll.), *PRL* **105** (2010) 252502.

[18] E.V. ShuryaK, *Phys. Rep.* **61** (1980) 71.

[19] P. Braun-Munzinger et al., http://arxiv.org/abs/nucl-th/0304013  april 3,2003.

[20] M.M. Aggarwal et al. (STAR Coll.), *PRL* **105** (2010) 022302.

[21] M. Kitazawa et al., *PRC* **86** (2012) 024904.

[22] L. Cifarelli et al., *Europhysicsnews*,**43-2** (2011) 29.

[23] G. Mouze and C. Ythier, http://arxiv.org/abs/1201.3819  [nucl-exp] Jan.19, 2012.

[24] H. Faust, 3rd Intern. Conf. on Dynamical Aspects of Nuclear Fission, Casta Papiernicka, Slovakia, J. Kliman and B.I. Pustylnik,eds., J.I.N.R., Dubna, 1996, p.63.

[25] M. Goeppert-Mayer and J.H.D. Jensen, Elementary Theory of Nuclear Shell Structures, New York, 1955.

[26] W. Heisenberg, *Z. Physik* **77** (1932) 1.

[27] K.F. Flynn et al., *PRC* **5** (1972) 1725.

[28] G.Mouze, S. Hachem and C. Ythier, http://arxiv.org/abs/1006.4068  [nucl-exp] June 29,2010.

[29] R.A. Ricci,  *Europhysicsnews*  **40-5** (2009) 13.

[30] Yu. V. Pyatkov et al*., EPJA* **45** (2010) 29.

[31] G. Mouze and C. Ythier, http://arxiv.org/abs/1202.1996  [nucl-exp] febr.10, 2012.

[32] G. Mouze, S. Hachem and C. Ythier, http://arxiv.org/abs/1101.1819   [nucl-exp] Jan.10,2011.